\documentclass[runningheads]{llncs}
\usepackage{graphicx}
\usepackage[english]{babel}
\usepackage{color, soul}
\usepackage{amsmath,amssymb,amsfonts}
\usepackage[colorlinks=true, linkcolor=blue, citecolor=blue, urlcolor=blue]{hyperref}

\begin{document}
\title{Modulation of viability signals for self-regulatory control}
%
%
\author{Alvaro Ovalle \and Simon M. Lucas}
\authorrunning{A. Ovalle and S. M. Lucas}
%
\institute{Queen Mary University of London. London, UK\\
\email{\{a.ovalle,simon.lucas\}@qmul.ac.uk}}
\maketitle              
\begin{abstract}

We revisit the role of instrumental value as a driver of adaptive behavior. In active inference, instrumental or extrinsic value is quantified by the information-theoretic \textit{surprisal} of a set of observations measuring the extent to which those observations conform to prior beliefs or preferences. That is, an agent is expected to seek the type of evidence that is consistent with its own model of the world. For reinforcement learning tasks, the distribution of preferences replaces the notion of reward. We explore a scenario in which the agent learns this distribution in a self-supervised manner. In particular, we highlight the distinction between observations induced by the environment and those pertaining more directly to the continuity of an agent in time. We evaluate our methodology in a dynamic environment with discrete time and actions. First with a surprisal minimizing model-free agent (in the RL sense) and then expanding to the model-based case to minimize the expected free energy.

\keywords{perception-action loop \and active inference \and reinforcement learning \and self-regulation \and anticipatory systems \and instrumental value.}
\end{abstract}
\section{Introduction}
The continual interaction that exists between an organism and the environment requires an active form of regulation of the mechanisms safeguarding its integrity. There are several aspects an agent must consider, ranging from assessing various sources of information to anticipating changes in its surroundings. In order to decide what to do, an agent must consider between different courses of action and factor in the potential costs and benefits derived from its hypothetical future behavior. This process of selection among different value-based choices can be formally described as an optimization problem. Depending on the formalism, the cost or utility functions optimized by the agent presuppose different normative interpretations. 

In reinforcement learning (RL) for instance, an agent has to maximize the expected reward guided by a signal provided externally by the environment in an oracular fashion. The reward in some cases is also complemented with an \textit{intrinsic} contribution, generally corresponding to an epistemic deficiency within the agent. For example prediction error \cite{pathak2017}, novelty \cite{NIPS2016_6383,burda2018,ostrovski2017} or ensemble disagreement \cite{pathak2019a}. It is important to note that incorporating these surrogate rewards into the objectives of an agent is often regarded as one of many possible enhancements to increase its performance, rather than been motivated by a concern with explaining the roots of goal-directed behavior.

In active inference \cite{friston2016b}, the optimization is framed in terms of the minimization of the \textit{variational free energy} to try to reduce the difference between sensations and predictions. Instead of rewards, the agent holds a prior over preferred future outcomes, thus an agent minimizing its free energy acts to maximize the occurrence of these preferences and to minimize its own \textit{surprisal}. Value arises not as an external property of the environment, but instead it is conferred by the agent as a contextual consequence of the interplay of its current configuration and the interpretation of stimuli.

There are recent studies that have successfully demonstrated how to reformulate RL and control tasks under the active inference framework. While for living processes it is reasonable to assume that the priors emerge and are refined over evolutionary scales and during a lifetime, translating this view into a detailed algorithmic characterization raises important considerations because there is no evolutionary prior to draw from. Thus the approaches to specify a distribution of preferences have included for instance, taking the reward an RL agent would receive and encoding it as the prior \cite{friston2012a,millidge2020a,sajid2020,tschantz2019,tschantz2020a,ueltzhoffer2018}, connecting it to task objectives \cite{sajid2020} or through expert demonstrations \cite{catal2019a,catal2020a,sancaktar2020}.

In principle this would suggest that much of the effort that goes into reward engineering in RL is relocated to that of specifying preferred outcomes or to the definition of a phase space. Nonetheless active inference provides important conceptual adjustments that could potentially facilitate conceiving more principled schemes towards a theory of agents that could provide a richer account of autonomous behavior and self-generation of goals, desires or preferences. These include the formulation of objectives and utilities under a common language residing in belief space, and appealing to a worldview in which utility is not treated as independent or detached from the agent. In particular the latter could encourage a more organismic perspective of the agent in terms of the perturbations it must endure and the behavioral policies it attains to maintain its integrity \cite{dipaolo2003}.

Here we explore this direction by considering how a signal acquires functional significance as the agent identifies it as a condition necessary for its viability and future continuity in the environment. Mandated by an imperative to minimize surprisal, the agent learns to associate sensorimotor events to specific outcomes. First, we start by introducing the surprise minimizing RL (SMiRL) specification \cite{berseth2020} before we proceed with a brief overview of the expected free energy. Then we motivate our approach from the perspective of a self-regulatory organism. Finally, we present results from our case study and close with some observations and further potential directions.

\section{Preliminaries}

\subsection{Model-free surprisal minimization} \label{dqn}

Consider an environment whose generative process produces a state $s_t \in \mathcal{S}$ at each time step $t$ resulting in an agent observing $o_t \in \mathcal{O}$. The agent acts on the environment with $a_t \in \mathcal{A}$ according to a policy $\pi$, obtaining the next observation $o_{t+1}$. Suppose the agent performs density estimation on the last $t-k$ observations to obtain a current set of parameter(s) $\theta_t$ summarizing $p_{\theta}(o)$. As these sufficient statistics contain information about the agent-environment coupling, they are concatenated with the observations into an augmented state $x_t = (o_t, \theta_t)$. Every time step, the agent computes the surprisal generated by a new observation given its current estimate and then updates it accordingly. In order to minimize surprisal under this model-free RL setting, the agent should maximize the expected log of the model evidence $\mathbb{E}[\sum_t \gamma^t \ln p_{\theta_{t}}(o_t)]$ \cite{berseth2020}. Alternatively, we maintain consistency with active inference by expressing the optimal surprisal Q-function as,

\begin{equation}
    Q_{\pi^*}(x, a) = \mathbb{E}_\pi[-\ln p_\theta(o) + \gamma \min_{a'} Q_{\pi^*}(x',a')]
\end{equation}
    
estimated via DQN \cite{mnih2013a} or any function approximator with parameters $\phi$ such that $Q_{\pi^*}(x,a) \approx Q(x,a;\phi)$.

\subsection{Expected free energy}

The free energy principle (FEP) \cite{friston2006} has evolved from an account of message passing in the brain to propose a probabilistic interpretation of self-organizing phenomena \cite{friston2013,ramstead2019,ramstead2018}. Central to current discourse around the FEP is the notion of the Markov blanket to describe a causal separation between the internal states of a system from external states, as well as the interfacing blanket states (i.e. sensory and active states). The FEP advances the view that a system remains far from equilibrium by maintaining a low entropy distribution over the states it occupies during its lifetime. Accordingly, the system attempts to minimize the surprisal of an event at a particular point in time.

This can be more concretely specified if we consider a distribution $p(o)$ encoding the states, drives or desires the system should fulfil. Thus the system strives to obtain an outcome $o$ that minimizes the surprisal $- \ln p(o)$. Alternatively, we can also state this as the agent maximizing its model evidence or marginal likelihood $p(o)$. For most cases estimating the actual marginal is intractable, therefore a system instead minimizes the free energy \cite{dayan1995,hinton1993a} which provides an upper bound on the log marginal \cite{jordan1999},

\begin{equation}\label{eq:fe}
\mathbf{F} = \mathbb{E}_{q(s)}[\ln q(s) -\ln p(o,s)]
\end{equation}

where $p(o,s)$ is the generative model and $q(s)$ the variational density approximating hidden causes. Equation \ref{eq:fe} is used to compute a static form of free energy and infer hidden causes given a set of observations. However if we instead consider an agent that acts over an extended temporal dimension, it must infer and select policies that minimize the expected free energy (EFE) $\mathbf{G}$ \cite{friston2016b} of a policy $\pi$ for a future step $\tau>t$. This can be expressed as,

\begin{equation}    
\mathbf{G}(\pi, \tau) = \mathbb{E}_{q(o_{\tau}, s_{\tau}|\pi)}[\ln q(s_\tau|\pi) - \ln p(o_\tau,s_\tau|\pi)]
\end{equation}

where $p(o_\tau,s_\tau|\pi)=q(s_\tau|o_\tau,\pi)p(o_\tau)$ is the generative model of the future. Rearranging $\mathbf{G}$ as,

\begin{equation}\label{eq:efe}
    \mathbf{G}(\pi, \tau) = - \underbrace{\mathbb{E}_{q(o_\tau|\pi)}[\ln p(o_\tau)]}_{instrumental\ value} -\underbrace{\mathbb{E}_{q(o_\tau|\pi)}\big[D_{KL}[\ln q(s_\tau|o_\tau,\pi)||\ln q(s_\tau|\pi)]\big]}_{epistemic\ value} 
\end{equation}

which illustrates how the EFE entails a pragmatic, instrumental or goal-seeking term that realizes preferences and an epistemic or information seeking term that resolves uncertainty. An agent selects a policy with probability $q(\pi) = \sigma(-\beta \sum_\tau \mathbf{G_\tau}(\pi))$ where $\sigma$ is the softmax function and $\beta$ is the inverse temperature. In summary, an agent minimizes its free energy via active inference by changing its beliefs about the world or by sampling the regions of the space that conforms to its beliefs.

\section{Adaptive control via self-regulation}

The concept of homeostasis has played a crucial role in our understanding of physiological regulation. It describes the capacity of a system to maintain its internal variables within certain bounds. Recent developments in the FEP describing the behavior of self-organizing systems under the framework, can be interpreted as an attempt to provide a formalization of this concept \cite{ramstead2018}. From this point of view, homeostatic control in an organism refers to the actions necessary to minimize the surprisal of the values reported by interoceptive channels, constraining them to those favored by a viable set of states. Something that is less well understood is how these attracting states come into existence. That is, how do they emerge from the particular conditions surrounding the system and how are they discovered among the potential space of signals. 

Recently, it has been shown that complex behavior may arise by minimizing surprisal in observation space (i.e. sensory states) without pre-encoded fixed prior distributions in large state spaces \cite{berseth2020}. Here we consider an alternative angle intended to remain closer to the homeostatic characterization of a system. In our scenario, we assume that given the particular dynamics of an environment, if an agent is equipped only with a basic density estimation capacity, then structuring its behavior around the type of regularities in observation space that can sustain it in time will be difficult. In these situations with fast changing dynamics, rather than minimizing free energy over sensory signals, the agent may instead leverage them to maintain a low future surprisal of another target variable. That implies that although the agent may have in principle access to multiple signals it might be interested in maintaining only some of them within certain expected range. 

Defining what should constitute the artificial physiology in simulated agents is not well established. Therefore we assume the introduction of an information channel representing in abstract terms the interoceptive signals that inform the agent about its continuity in the environment. We can draw a rudimentary comparison, and think of this value in a similar way in which feelings agglutinate and coarse-grain the changes of several internal physical responses \cite{damasio2004}. In addition, we are interested in the agent learning to determine whether it is conductive to its self-preservation in the environment or not.

\subsection{Case Study} 

We assess the behavior of an agent in the \textit{Flappy Bird} environment (fig. \ref{fig:results} left). This is a task where a bird must navigate between obstacles (pipes) at different positions while stabilizing its flight. Despite the apparent simplicity, the environment offers a fundamental aspect present in the physical world. Namely, the inherent dynamics leads spontaneously to the functional disintegration of the agent. If the agent stops propelling, it succumbs to gravity and falls. At the same time the environment has a constant scrolling rate, which implies that the agent cannot remain floating at a single point and cannot survive simply by flying aimlessly. Originally, the task provides a reward every time the bird traverses in between two pipes. However for our case study the information about the rewards is never propagated and therefore does not have any impact on the behavior of the agent. The agent receives a feature vector of observations indicating its location and those of the obstacles. In addition, the agent obtains a \textit{measurement} $v$ indicating its presence in the task (i.e. 1 or 0). This measurement does not represent anything positive or negative by itself, it is simply another signal that we assume the agent is able to calculate. Similarly to the outline in \ref{dqn}, the agent monitors the last $t-k$ values of this measurement and estimates the density to obtain $\theta_t$. These become the statistics describing the current approximated distribution of preferences $p(v|\theta_t)$ or $p_{\theta_t}(v)$, which are also used to augment the observations to $x_t=(o_t,\theta_t)$. When the agent takes a new measurement $v_{t}$, it evaluates the surprisal against $p_{\theta_{t-1}}(v_t)$. In this particular case it is evaluated via a Bernoulli density function such that $-\ln p_{\theta_{t-1}}(v_t) = - (v_t \ln \theta_{t-1} + (1-v_t) \ln (1-\theta_{t-1}))$. First, we train a baseline model-free surprisal minimizing DQN as specified in \ref{dqn} parameterized by a neural network (NN). Then we examine the behavior of a second agent that minimizes the expected free energy. Thus the agent learns an augmented state transition model of the world, parameterized by an ensemble of NNs, and an expected surprisal model, also parameterized by another NN. In order to identify an optimal policy we apply rolling horizon evolution \cite{perez2013} to generate candidate policies $\pi=(a_\tau,...,a_T)$ and to associate them to an expected free energy given by (appendix \ref{ap:a}),

\begin{equation}\label{eq:hefe}
    \mathbf{G}(\pi,\tau) \approx -\mathbb{E}_{q(o_\tau,v_\tau,\theta|\pi)}D_{KL}[q(s_\tau|,o_\tau,v_\tau,\pi)||q(s_\tau|\pi)] - \mathbb{E}_{q(v_\tau,\theta,s_\tau|\pi)}[\ln p_\theta(v_\tau)]
\end{equation}

If we explicitly consider the model parameters $\phi$, equation \ref{eq:hefe} can be decomposed as (appendix \ref{ap:b}),

\begin{align}\label{eq:computable}
    \mathbf{G}(\pi,\tau) &\approx -\underbrace{\mathbb{E}_{q(o_\tau,v_\tau,\phi|\pi)}D_{KL}[q(s_\tau|o_\tau,v_\tau,\pi)||q(s_\tau|\pi)]}_{salience}\nonumber\\ 
    &\quad -\underbrace{\mathbb{E}_{q(o_\tau,v_\tau,s_\tau|\pi)}D_{KL}[q(\phi|s_\tau, o_\tau, v_\tau, \pi)||q(\phi)]}_{novelty}\nonumber\\ 
    &\quad - \underbrace{\mathbb{E}_{q(o_\tau,v_\tau,s_\tau,\phi|\pi)}[\ln p_\theta(v_\tau)]}_{instrumental\ value}\nonumber\\
\end{align}

The expression unpacks further the epistemic contributions to the EFE in terms of salience and novelty \cite{friston2017}. These terms refer to the expected reduction in uncertainty about hidden causes and in the parameters respectively. For this task $o=s$, thus only the first and third term are considered.

\subsection{Evaluation}

\begin{figure}[t]
    \includegraphics[scale=0.12]{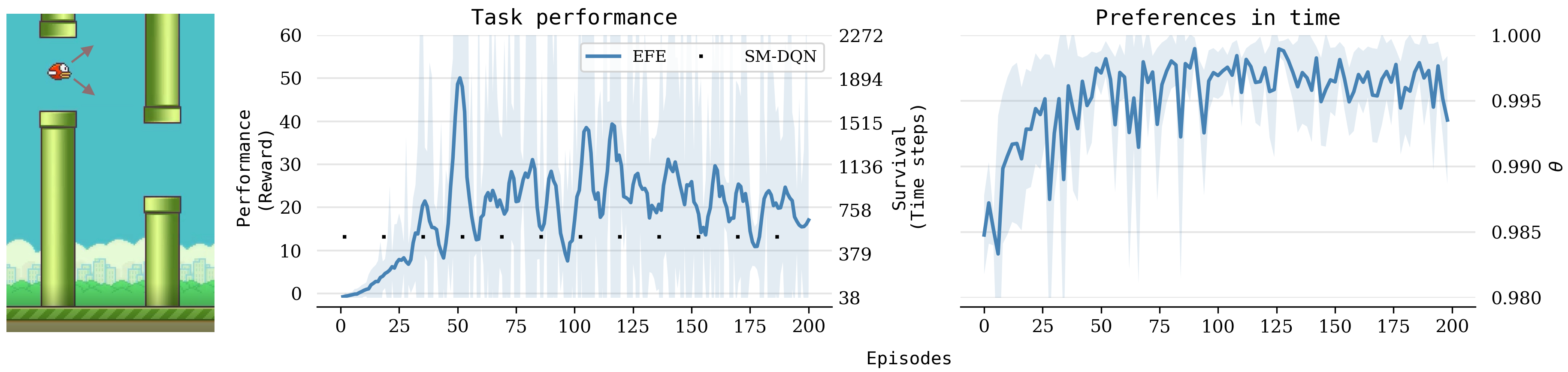}
    \centering
    \caption{\textit{Left}: The Flappy Bird environment. \textit{Center}: Performance of an EFE agent. The left axis indicates the unobserved rewards as reported by the task and the right axis the number of time steps it survives in the environment. The dotted line shows the average performance of an SM-DQN after 1000 episodes. \textit{Right}: Parameter $\theta$ in time, summarizing the intra-episode sufficient statistics of $p_\theta(v)$. } \label{fig:results}
\end{figure}

The plot on fig. \ref{fig:results} (center) tracks the performance of an EFE agent in the environment (averaged over 10 seeds). The dotted line represents the surprisal minimizing DQN agent after 1000 episodes. The left axis corresponds to the (unobserved) task reward while the right axis indicates the approximated number of time steps the agent survives. During the first trials, and before the agent exhibits any form of competence, it was observed that the natural coupling between agent and environment grants the agent a life expectancy of roughly 19-62 time steps in the task. This is essential as it starts to populate the statistics of $v$.  Measuring a specific quantity $v$, although initially representing just another signal, begins to acquire certain value due to the frequency that it occurs. In turn, this starts to dictate the preferences of the agent as it hints that measuring certain signal correlates with having a stable configuration for this particular environment as implied by its low surprisal. Right fig. \ref{fig:results} shows the evolution of parameter $\theta$ (averaged within an episode) corresponding to the distribution of preferred measurements $p_\theta(v)$ which determines the level of surprisal assigned when receiving the next $v$. As the agent reduces its uncertainty about the environment it also becomes more capable of associating sensorimotor events to specific measurements. The behavior becomes more consistent with seeking less surprising measurements, and as we observe, this reinforces its preferences, exhibiting the circular self-evidencing dynamics that characterize an agent minimizing its free energy.

\section{Discussion}

\textbf{Learning preferences in active inference:}
The major thesis in active inference is the notion of an agent acting in order to minimize its expected surprise. This implies that the agent will exhibit a tendency to seek for the sort of outcomes that have high prior probability according to a biased model of the world, giving rise to goal-directed behavior. Due to the difficulty of modeling an agent to exhibit increasing levels of autonomy, the agent based simulations under this framework, and similarly to how it has largely occurred in RL, have tended to concentrate on the generation of a particular expected behavior in the agent. That is, on how to make the agent perform a task by encoding predefined goals \cite{friston2012a,millidge2020a,sajid2020,tschantz2019,tschantz2020a,ueltzhoffer2018} or providing guidance \cite{catal2019a,catal2020a,sancaktar2020}. However there has been recent progress trying to mitigate this issue. For example, in some of the simulations in \cite{sajid2020} the authors included a distribution over prior preferences to account for each of the cells in \textit{Frozen Lake}, a gridworld like environment. Over time the prior preferences are tuned, leading to habit formation. Most related to our work, are the studies on surprise minimizing RL (SMiRL) by \cite{berseth2020}, where model-free agents performed density estimation on their observation space and acquired complex behavior in various tasks by maximizing the model evidence of their observations. Here we have also opted for this approach, however we have grounded it on organismic based considerations of viability as inspired by insights on the nature of agency and adaptive behavior \cite{barandiaran2009,dipaolo2010,dipaolo2003}. It has been suggested that even if some of these aspects are defined exogenously they could capture general components of all physical systems and could potentially be derived in a more objective manner compared to task based utilities \cite{kolchinsky2018a}. Moreover these views suggest that the inherent conditions of precariousness and the perturbations an agent must face are crucial ingredients for the emergence of purpose generating mechanisms. In that sense, our main concern has been to explore an instance of the conditions in which a stable set of attracting states arises, conferring value to observations and leading to what seemed as self-sustaining dynamics. Although all measurements lacked any initial functional value, the model presupposes the capacity of the agent to measure its operational integrity as it would occur in an organism monitoring its bodily states. This raises the issue of establishing more principled protocols to define what should constitute the internal milieu of an agent.

\noindent\textbf{Agent-Environment coupling:} A matter of further analysis, also motivated by results in \cite{berseth2020}, is the role of the environment to provide structure to the behavior of the agent. For instance, in the environments in \cite{berseth2020}, a distribution of preferences spontaneously built on the initial set of visual observations tends to correlate with good performance on the task. In the work presented here the initial set of internal measurements afforded by the environment contributes to the formation of a steady state, with the visual features informing the actions necessary to maintain it. Hence similarly to \cite{berseth2020}, the initial conditions of the agent-environment coupling that furnish the distribution $p(v)$ provide a starting solution for the problem of self-maintenance as long as the agent is able to preserve the statistics. Thus if the agent lacks a sophisticated sensory apparatus, the capacity to extract invariances or the initial statistics of sensory data do not favor the emergence of goal-seeking behavior, tracking its internal configuration may suffice for some situations. However this requires further unpacking, not only because as discussed earlier it remains uncertain how to define the internal aspects of an agent, but also because often simulations do not capture the essential characteristics of real environments either \cite{co-reyes2020}. 

\noindent\textbf{Drive decomposition:} While here we have afforded our model certain levels of independence between the sensory data and the internal measurements, it might be sensible to imagine that internal states would affect perception and perceptual misrepresentation would affect internal states. Moreover, as the agent moves from normative conditions based entirely on viability to acquire other higher level preferences, it learns to integrate and balance different drives and goals. From equation \ref{eq:ap} it is also possible to conceive a simplified scenario and establish the following expression (appendix \ref{ap:d}),

\begin{align}
\mathbf{G}(\pi,\tau)&\approx \underbrace{\mathbb{E}_{q(o_\tau,v_\tau,\theta,s_\tau|\pi)}[\ln q(s_\tau|\pi) - \ln p(s_\tau|o_\tau,\pi)]}_{epistemic\ value}\nonumber\\ 
&\quad - \underbrace{\mathbb{E}_{q(o_\tau,v_\tau,\theta,s_\tau|\pi)}[\ln p(o_\tau)]}_{high\ level\ value}\nonumber\\
    &\quad + \underbrace{\mathbb{E}_{q(o_\tau, s_\tau|\pi)} H[p(v_\tau|s_\tau,o_\tau,\pi)]}_{regulatory\ value}\nonumber\\
\end{align}

Where the goal-seeking value is decomposed into a component that considers preferences encoded in a distribution $p(o)$ and another element estimating the expected entropy of the distribution of essential variables. Policies would balance the contributions resolving for hypothetical situations, such as a higher level goal being at odds with the viability of the system. 

\section*{Acknowledgment}

This research utilised Queen Mary's Apocrita HPC facility, supported by QMUL Research-IT. doi:10.5281/zenodo.438045

\bibliographystyle{splncs04}
\bibliography{refs}

\appendix

\section{Expected Free Energy with measurements $v$}\label{ap:a}

We consider a generative model $p(s,o,v|\pi)$ for the EFE equation and obtain a joint distribution of preferences $p(o,v)$. If we are interested exclusively in $v$, assuming and treating $o$ and $v$ as if they were independent modalities, and ignoring $o$ we obtain:

\begin{align}
    \mathbf{G}(\pi,\tau) &= \mathbb{E}_{q(o_\tau,v_\tau,\theta,s_\tau|\pi)}[\ln q(s_\tau|\pi) - \ln p(s_\tau,o_\tau,v_\tau|\pi)]\label{eq:ap}\\
    &\approx \mathbb{E}_{q(o_\tau,v_\tau,\theta,s_\tau|\pi)}[\ln q(s_\tau|\pi) - \ln q(s_\tau|,o_\tau,v_\tau,\pi) - \ln p(o_\tau,v_\tau)]\nonumber\\
    &\approx \mathbb{E}_{q(o_\tau,v_\tau,\theta,s_\tau|\pi)}[\ln q(s_\tau|\pi) - \ln q(s_\tau|,o_\tau,v_\tau,\pi) - \ln p(o_\tau) - \ln p_\theta(v_\tau)]\nonumber\\
    &\approx \mathbb{E}_{q(o_\tau,v_\tau,\theta,s_\tau|\pi)}[\ln q(s_\tau|\pi) - \ln q(s_\tau|,o_\tau,v_\tau,\pi) - \ln p_\theta(v_\tau)]\nonumber\\
    &\approx -\mathbb{E}_{q(o_\tau,v_\tau,\theta|\pi)}D_{KL}[q(s_\tau|,o_\tau,v_\tau,\pi)||q(s_\tau|\pi)] - \mathbb{E}_{q(v_\tau,\theta,s_\tau|\pi)}[\ln p_\theta(v_\tau)]\nonumber\\
\end{align}

\section{Novelty and salience}\label{ap:b}

The derivation is equivalent to those found in the classical tabular descriptions of active inference where instead of learning transitions via a function approximator, a mapping from hidden states to observations is encoded by a likelihood matrix $\mathbf{A}$. In the tabular case the beliefs of the probability of an observation given a state are contained in the parameters $a_{ij}$, which are updated as the agent obtains a particular observation.

\begin{align}
    \mathbf{G}(\pi,\tau) &= \mathbb{E}_{q(o_\tau, s_\tau, v_\tau,\phi|\pi)} [\ln q(s_\tau,\phi|\pi) - \ln p(o_\tau,v_\tau,s_\tau,\phi|\pi)]\nonumber\\
    &= \mathbb{E}_{q(o_\tau, s_\tau, v_\tau, \phi|\pi)} [\ln q(\phi) + \ln q(s_\tau|\pi\nonumber)\\
    &\quad - \ln p(\phi|s_\tau, o_\tau, v_\tau, \pi) - \ln p(s_\tau|o_\tau,v_\tau,\pi) - \ln p(o_\tau,v_\tau)]\nonumber\\
&\approx \mathbb{E}_{q(o_\tau, s_\tau, v_\tau, \phi|\pi)} [\ln q(\phi) + \ln q(s_\tau|\pi)\nonumber\\
&\quad - \ln q(\phi|s_\tau, o_\tau, v_\tau,\pi) - \ln q(s_\tau|o_\tau,v_\tau,\pi) - \ln p_\theta(v_\tau)]\nonumber\\
&\approx \mathbb{E}_{q(o_\tau, s_\tau, v_\tau, \phi|\pi)} [\ln q(s_\tau|\pi) - \ln q(s_\tau|o_\tau,v_\tau,\pi)]\nonumber\\
&\quad + \mathbb{E}_{q(o_\tau, s_\tau, v_\tau \phi|\pi)} [\ln q(\phi)- \ln q(\phi|s_\tau, o_\tau, v_\tau,\pi)] \nonumber\\
&\quad - \mathbb{E}_{q(o_\tau, s_\tau, v_\tau, \phi|\pi)} [\ln p(v_\tau)]\nonumber\\
&\approx - \mathbb{E}_{q(o_\tau, s_\tau, v_\tau, \phi|\pi)} [\ln q(s_\tau|o_\tau,v_\tau,\pi)-\ln q(s_\tau|\pi)]\nonumber\\
&\quad -\mathbb{E}_{q(o_\tau, s_\tau, v_\tau, \phi|\pi)} [\ln q(\phi|s_\tau, o_\tau, v_\tau, \pi) - \ln q(\phi)]\nonumber\\
&\quad - \mathbb{E}_{q(o_\tau, s_\tau, v_\tau, \phi|\pi)} [\ln p(v_\tau)]\nonumber\\
&\approx - \underbrace{\mathbb{E}_{q(o_\tau, v_\tau, \phi|\pi)} \big[D_{KL}[q(s_\tau|o_\tau,v_\tau,\pi)||q(s_\tau|\pi)]\big]}_{salience}\nonumber\\
&\quad -\underbrace{\mathbb{E}_{q(o_\tau, v_\tau, s_\tau|\pi)} \big[D_{KL}[q(\phi|s_\tau, o_\tau, v_\tau, \pi)|| q(\phi)]\big]}_{novelty}\nonumber\\
&\quad - \underbrace{\mathbb{E}_{q(o_\tau, v_\tau, s_\tau, \phi|\pi)} [\ln p(v_\tau)]}_{instrumental\ value}\nonumber\\
\end{align}

\section{Implementation}\label{ap:c}

We tested on the Flappy Bird environment \cite{tasfi2016}. The environment sends a non-visual vector of features containing:

\begin{itemize}
\item the bird $y$ position
\item the bird velocity.
\item next pipe distance to the bird
\item next pipe top $y$ position
\item next pipe bottom $y$ position
\item next next pipe distance to the bird
\item next next pipe top $y$ position
\item next next pipe bottom $y$ position
\end{itemize}

The parameter $\theta$ of the Bernoulli distribution $p(v)$ was estimated from a \textit{measurement buffer} (i.e. queue) containing the last $N$ values of $v$ gathered by the agent. We tested the agents with large buffers (e.g. $20^6$) as well as small buffers (e.g. $20$) without significant change in performance. The results reported in fig. \ref{fig:results} were obtained with small sized buffers as displayed in the hyperparameter table below.

The DQN agent was trained to approximate with a neural network a Q-function $Q_\phi(\{s,\theta\},.)$. For our case study $s=o$ which contains the vector of features, while $\theta$ is the parameter corresponding to the current estimated statistics of $p(v)$. An action is sampled uniformly with probability $\epsilon$ otherwise $a_t =\min_a Q_\phi(\{s_t,\theta_t\},a)$. $\epsilon$ decays during training.

For the EFE agent, the transition model $p(s_t|s_{t-1},\phi,\pi)$ is implemented as a $\mathcal{N} (\{s_t,\theta_t\}; f_\phi(s_{t-1},\theta_{t-1},a_{t-1}), f_\phi(s_{t-1}, \theta_{t-1}, a_{t-1}))$.  Where $a$ is an action of a current policy $\pi$ with one-hot encoding and $f_\phi$ is an ensemble of $K$ neural networks which predicts the next values of $s$ and $\theta$. The surprisal model is also implemented with a neural network and trained to predict directly the surprisal in the future as $f_\xi(s_{t-1},\theta_{t-1},a_{t-1})=-\ln p_{\theta_{t-1}}(v_t)$.

In order to calculate the expected free energy in equation \ref{eq:computable} from a simulated sequence of future steps, we follow the approach described in appendix G in \cite{tschantz2020a} where they show that an information gain of the form $\mathbb{E}_{q(s|\phi)}D_{KL}[q(\phi|s)||q(\phi)]$ can be decomposed as,

\begin{equation}
\mathbb{E}_{q(s|\phi)}D_{KL}[q(\phi|s)||q(\phi)]=-\mathbb{E}_q(\phi)H[q(s|\phi)] + H[\mathbb{E}_{q(\phi)}q(s|\phi)]
\end{equation}

with the first term computed analytically from the ensemble output and the second term approximated with a k-NN estimator \cite{beirlant1997}.

\begin{center}
    \begin{tabular}{||c c c||} 
    \hline
    Hyperparameters & DQN & EFE \\ [0.5ex] 
    \hline\hline
    Measurement $v$ buffer size &  20 & 20 \\ 
    \hline
    Replay buffer size &  $10^6$ & $10^6$ \\ 
    \hline
    Batch size & 64 & 50 \\ 
    \hline
    Learning rate & $1^{-3}$ & $1^{-3}$ \\
    \hline
    Discount rate & 0.99 & - \\ 
    \hline
    Final $\epsilon$ & 0.01 & - \\ 
    \hline
    Seed episodes & 5 & 3 \\
    \hline
    Ensemble size & - & 25 \\
    \hline
    Planning horizon & - & 15 \\ 
    \hline
    Number of candidates & - & 500 \\ 
    \hline
    Mutation rate & - & 0.5 \\
    \hline
    Shift buffer & - & True \\ 
    \hline
   \end{tabular}
   \end{center}

\section{Drive decomposition}\label{ap:d}

\begin{align}
    \mathbf{G}(\pi,\tau) &= \mathbb{E}_{q(o_\tau,v_\tau,\theta,s_\tau|\pi)}[\ln q(s_\tau|\pi) - \ln p(s_\tau,o_\tau,v_\tau|\pi)]\nonumber\\
    &= \mathbb{E}_{q(o_\tau,v_\tau,\theta,s_\tau|\pi)}[\ln q(s_\tau|\pi) - \ln p(v_\tau|s_\tau,o_\tau,\pi) - \ln p(s_\tau, o_\tau|\pi)]\nonumber\\
    &= \mathbb{E}_{q(o_\tau,v_\tau,\theta,s_\tau|\pi)}[\ln q(s_\tau|\pi) - \ln p(s_\tau|o_\tau,\pi) -\ln p(o_\tau) - \ln p(v_\tau|s_\tau,o_\tau,\pi)]\nonumber\\
    &\approx \mathbb{E}_{q(o_\tau,v_\tau,\theta,s_\tau|\pi)}[\ln q(s_\tau|\pi) - \ln p(s_\tau|o_\tau,\pi)] - \mathbb{E}_{q(o_\tau,v_\tau,\theta,s_\tau|\pi)}[\ln p(o_\tau)]\nonumber\\
    &\quad + \mathbb{E}_{q(o_\tau, s_\tau|\pi)} H[p(v_\tau|s_\tau,o_\tau,\pi)]\nonumber\\
\end{align}
\end{document}